# Flexo-diffusion effect: the strong influence on lithium diffusion induced by strain gradient

Gao Xu,[a] Feng Hao,[b] Mouyi Weng,[c] Jiawang Hong,[*d] Feng Pan,[*c] and Daining Fang[*ae]

[a]*State Key Laboratory for Turbulence and Complex Systems & Center for Applied Physics and Technology, College of Engineering, Peking University, Beijing 100871, P.R. China*

[b]*Department of Engineering Mechanics, Shandong University, Jinan 250100, P.R. China*

[c]*School of Advanced Materials, Peking University, Shenzhen Graduate School, Shenzhen 518055, P.R. China.*

[d]*School of Aerospace Engineering, Beijing Institute of Technology, Beijing 100081, P.R. China*

[e]*Institute of Advanced Structure Technology, Beijing Institute of Technology, Beijing 100081, P.R. China*

[*]Corresponding author

*E-mail address*: hongjw@bit.edu.cn, panfeng@pkusz.edu.cn, fangdn@bit.edu.cn

## Abstract:

Lithium ion batteries (LIBs) work under sophisticated external force field and its electrochemical properties could be modulated by strain. Owing to the electro-mechanical coupling, the change of micro-local-structures can greatly affect lithium (Li) diffusion rate in solid state electrolytes and electrode materials of LIBs. In this study, we find that strain gradient in bilayer graphene (BLG) significantly affects Li diffusion barrier, which is termed as the flexo-diffusion effect, through first-principles calculations. The Li diffusion barrier substantially decreases/increases under the positive/negative strain gradient, leading to the change of Li diffusion coefficient in several orders of magnitude at 300 K. Interestingly, the regulation effect of strain gradient is much more significant than that of uniform strain field, which can have a remarkable effect on the rate performance of batteries, with a considerable increase in the ionic conductivity and a slight change of the original material structure. Moreover, our *ab initio* molecular dynamics simulations (AIMD) show that the asymmetric distorted lattice structure provides a driving force for Li diffusion, resulting in oriented diffusion along the positive strain gradient direction. These findings could extend present LIBs technologies by introducing the novel strain gradient engineering.

# 1. Introduction

Lithium (Li) ion batteries (LIBs) with high capacity and high rate have attracted tremendous attention. In order to increase the power density and charge/discharge efficiency of LIBs, it is imperative to improve the rate performance of electrode materials. From the fundamental physics point of view, it is essential to improve the electronic and ionic conductivity of the electrode materials.[1] Li-ion transport pathways and rate in LIBs have been widely studied by the first-principles calculations, e.g. the faster Li-ion diffusion vs. the larger layer distance in layered Li transition metal (TM) oxides $LiTMO_2$ (TM = Ni, Mn, Co, or $Ni_xMn_yCo_z$, $x+y+z=1$) as cathode materials,[2] and high-throughput methods.[3] New transport mechanisms, such as cooperative transport and liquid-like transport enabling fast Li-ion diffusion in solid electrolyte, were proposed,[4-6] and new materials were predicted under common conditions by focusing on materials itself.[7,8]

Previous successful strategies to improve diffusion properties were mainly related to chemical doping and morphology modulation, such as nanosizing and carbon coating.[9,10] Besides, strain engineering is also an effective approach to modulate the LIBs performance without changing the chemical constituents of electrode materials. Strain has been shown as an effective method for tuning electronic, transport, and optical properties of LIBs for decades.[11-16] In particular, the ionic conductivity of many electrode materials can be significantly enhanced by applying uniform strain.[17-21] A previous study showed that activation barrier is very sensitive to the Li-concentration due to the strongly varying *c*-lattice parameter of the layered intercalation compounds,[22] the essential reason for that is the uniform strain in *c*, i.e. out-of-plane, direction. With the rapid development of flexible electronics, the LIBs have also been considered as key components in flexible, stretchable and wearable devices,[23-25] which inevitably involve the strain gradient. Therefore, in addition to strain, strain gradient may introduce new regulative phenomenon and efficiently regulative effect to Li-ion diffusion in electrode materials. However, the strain gradient effect is rarely studied in LIBs.

Interestingly, the strain gradient effect was recently attracted tremendous attention in electro-mechanical materials,[26,27] termed as flexoelectricity which reflects a novel coupling between an electric polarization and a strain gradient.[28,29] Flexoelectricity is a universal property allowed by symmetry in any dielectric materials and therefore significantly broadens the choice of materials for the applications in electromechanical sensors and actuators. In addition, the magnitude of strain gradient in nanostructures (~ $10^9$ m$^{-1}$)[30,31] was found to be remarkably higher than that in bulk materials (~$10^0$ m$^{-1}$),[32,33] indicating it may significantly affect the properties at nanoscale or even induce new phenomena.[34-36] However, strain gradient has not been applied to modulate the diffusion properties in LIBs yet. Compared to the uniform strain modulation effect, strain gradient may also provide remarkable modulation effect or account for some novel phenomena for LIBs. For example, very recently, it was found that a giant strain gradient appears in deposited Li film on Cu substrate, which may drive the Li dendrite growth.[37] In this study, we will propose and demonstrate that the diffusion barrier, which dominates the ion conductivity, can be modulated by strain gradient significantly. Herein, the change of diffusion and its corresponding properties induced by strain gradient are termed as the flexo-diffusion effect.

To demonstrate this, we choose bilayer graphene (BLG) as prototype to explore the flexo-diffusion effect induced by the strain gradient. Graphene, the most well-studied two-dimensional (2D) material, exhibits unique capacity in LIBs because of its high charge carrier mobility,[38] large surface area[39] and a broad electrochemical window.[40] Up to date, graphene-based materials have been utilized as electrode materials for batteries with great success.[41-45] However, the value of Li diffusion barrier is 0.327 eV for graphene,[46] which is larger than that of other 2D materials, such as 0.25eV on $MoS_2$[47] and 0.08eV on phosphorene.[48] Furthermore, Li diffusion in carbonaceous nanostructures is still not understood fully because of the lack of reliable experimental methods.[49] Therefore, it is essential to theoretically evaluate the diffusion barrier and path of Li.[50] Strain is particularly useful when engineering 2D crystals because these reduced-dimensional structures can sustain much larger strains than bulk crystals. Monolayer graphene was reported to be strained up to its intrinsic limit (~25 %) without substantially damaging its crystal structures,[51] providing a dramatically wide range for tuning its mechanical and electronic performance. Recently, Chen et al.[52] reported that origami is an efficient way to convert graphene into atomically precise, complex nanostructures, which could controllably apply strain gradient in graphene at nanoscale. As a representative 2D layered electrode material, investigation of the flexo-diffusion effect of graphene can also provide a good reference for other 2D layered electrode materials and bulk crystals with imhomogenously distorted microstructures.

In this work, we investigate the strain gradient dependence of Li diffusion barrier in BLG from the first-principles method. We find the flexo-diffusion effect strongly influences the Li diffusion in BLG. It shows that the Li diffusion barrier decreases/increases substantially under the positive/negative strain gradient, and the ionic conductivity increases by several orders of magnitude with the positive strain gradient at 300 K. Besides, the modulation of strain gradient is more efficient than that of uniform strain. Moreover, the asymmetric Li potential energy distribution induced by strain gradient provides a driving force to drive Li oriented motion along the positive gradient direction, which can't be realized in uniform strain field. The flexo-diffusion effect provides a new approach to tune and optimize the lithium diffusion in LIBs applications.

## 2. Computational Details

All the calculations are performed using the Vienna *ab initio* simulation package (VASP),[53,54] which is based on the density functional theory (DFT). Projector-augmented-wave (PAW) potentials[55] are used to take into account the electron−ion interactions, while the electron exchange-correlation interactions are treated using generalized gradient approximation (GGA)[56] in the scheme of Perdew-Burke-Ernzerhof. Taking into consideration of the van der Waal forces, the results are calculated at the DFT-D2 level.[57] To analyze the Li diffusion in between layers of graphite, we simulate the Li diffusion in between two layers for BLG. The unit cell (8 carbon atoms) is obtained from AB stacked hexagonal graphite; the two orthogonal in-plane lattice vectors are used for subsequent applying strain gradient. Atomic relaxation of the unit cell is performed with the energy and force convergence criterion of $1 \times 10^{-6}$ eV and 0.001 eV/Å, respectively. A vacuum space of ~ 20 Å is placed between adjacent layers to avoid mirror interactions. A plane wave cutoff of 600 eV is set in our calculations with K-point samplings of $9 \times 12 \times 1$ grids. The carbon atom positions are fixed in other calculations. The diffusion barrier calculations and potential energy distribution calculations are performed on the $3 \times 2 \times 1$ and $4 \times 2 \times 1$ supercells, respectively. To simulate Li diffusion inside BLG, we perform minimum energy path profiling using the climbing image nudged elastic band (CI-NEB) method as implemented in the VASP transition state tools,[58,59] where Li positions are relaxed. The calculated diffusion barriers are defined in Supplementary Note 1. The kinetic properties of Li in different strain fields of BLG are studied using the *ab initio* molecular dynamics (AIMD) method. The simulations are performed on the $3 \times 2 \times 1$ supercell with only Γ point sampling in the Brillouin zone. We perform constant volume and temperature (NVT) AIMD calculations with a Nose thermostat[60] for 6000 steps with a time step of 1 fs.

## 3. Results and Discussion

Our calculated interlayer spacing of BLG is 3.24 Å, which agrees well with other theoretical calculations (3.25 Å[61] and 3.28 Å[62]) and is slightly smaller than experimental data for lattice parameters *c*/2 of bulk graphite (3.35 Å[63]). The pristine interlayer spacing of BLG is set to 3.24 Å, i.e. no applied strain. To analyze Li diffusion properties in BLG, the Li adsorption positions should be first verified. Fig. 1(a) shows the local configure of BLG, we follow a previous study[64] and consider two sites with high symmetry: the site on the top of a carbon atom (TT) and the site in the center of a hexagon (HT). For the stage-I compound $LiC_6$ of graphite, the HT is the most stable adsorption site.[64] To simulate the Li diffusion in between two layers of BLG, two high symmetrical paths of Li diffusion are considered, including HT-TT and HT-HT, as shown in Fig. 1(a). The results in Fig. 1(b) and 1(c) show that the path HT-HT has a length of 1.47 Å, with a much lower energy barrier (287 meV) than that of HT-TT diffusion path. Thus, the process of Li diffusion in BLG is repeatedly from one HT position to its adjacent HT position.

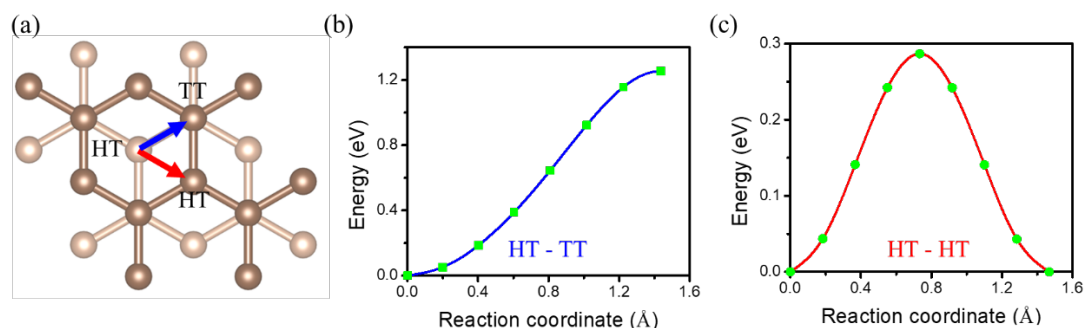

Fig. 1 (a) The schematic illustration for two possible diffusion paths in BLG. The dark brown and light brown represent the top and bottom carbon atomic layers, respectively. The HT and TT indicate two symmetry Li adsorption sites between BLG. The energy profiles for (b) path HT-TT and (c) path HT-HT.

To analyze the strain modulation effect on Li diffusion inside BLG, we explore Li diffusion inside the BLG in different strain fields. Fig. 2 shows the schematic calculation configure. Here, all the intercalated Li atoms are located in the left part of the supercell, as shown in Fig 2(a). The adsorption sites of Li labeled A to D in Fig. 2(a) are the main adsorption positions in the following calculations. We construct a supercell with periodic strain to induce strain gradient without breaking periodic boundary condition for DFT calculations,[65] as shown in Fig. 2(c) and 2(d). In this supercell, the bottom graphene layer keeps flat while the interlayer spacing $d(x)$ linearly changes following $d(x) = d_0 \times (1 + \eta x)$ until the middle of supercell ($x = L/2$, $L$ is the length of supercell in $x$ direction), where $d_0$ is the interlayer spacing of pristine BLG (3.24 Å) and $\eta$ is the constant applied strain gradient; then the interlayer spacing reversely changes from the middle to the end in the direction. The magnitude of strain in the middle of the $x$-lattice of supercell is defined as mid-strain $\varepsilon_m$ ($\varepsilon_m = \eta L/2$), which can be used to represent the strain gradient in an alternative way. Mid-strain is the maximum/minimum strain in the BLG with strain gradient. We adopt mid-strain $\varepsilon_m$ to compare with the uniform strain cases in which the strain magnitude is the same as $\varepsilon_m$. In this way, the difference between strain gradient case and uniform strain case will come from the strain gradient effect. We also define the positive strain gradient as $\eta > 0$ ($\varepsilon_m > 0$) and negative strain gradient as $\eta < 0$ ($\varepsilon_m < 0$), as shown in Fig. 2(c) and 2(d).

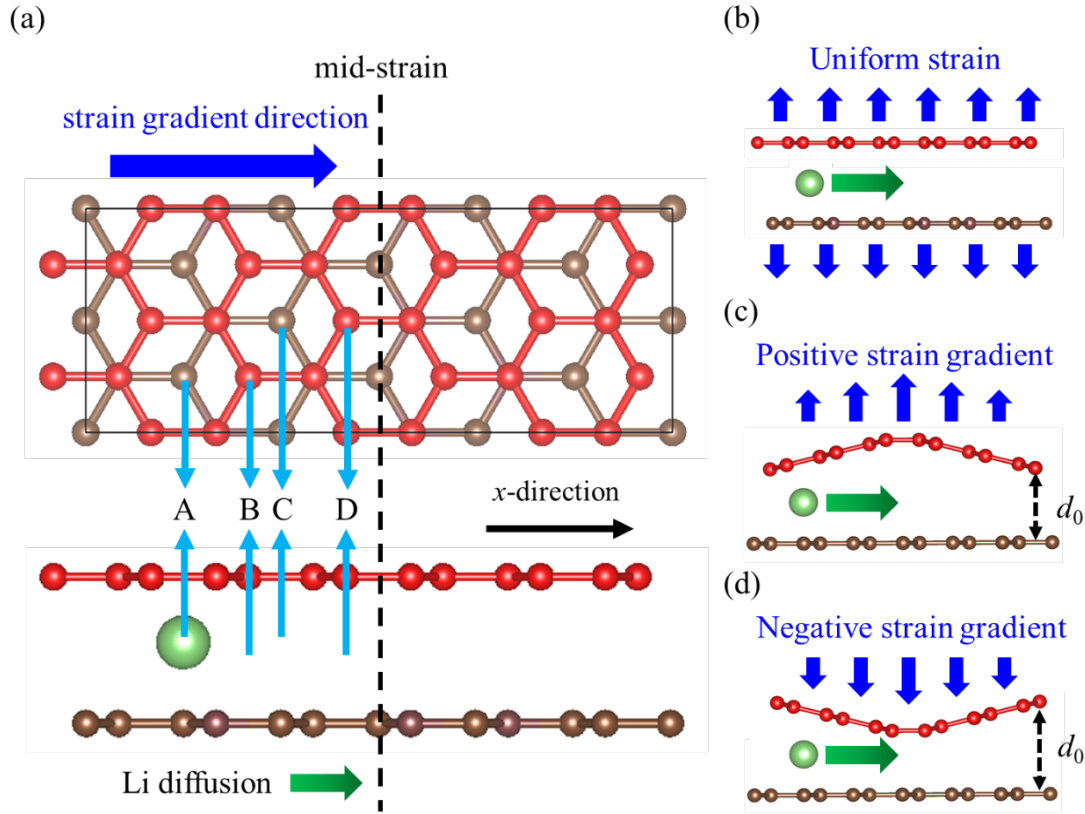

Fig. 2 (a) The top and side views of the pristine 3 × 2 × 1 supercell. The positions labeled A to D are the four stable HT sites for intercalated Li atoms. The schematic configures from side view for Li diffusion inside BLG: (b) uniform strain field, (c) positive strain gradient field, and (d) negative strain gradient field. All the applied strains are in the out-of-plane direction. In the sake of a visible applied strain gradient, the crystal configures in strain gradient field in this figure and the following figures are schematic configures. The red and brown represent the top and bottom graphene layers, and the black frame indicates the supercell.

Then, we first consider the effect of uniform strain on Li diffusion in BLG. Uniaxial stress is applied in the out-of-plane direction. The energy profiles for the Li pathways are calculated in four cases: pristine (0 %), ±10 % and 20 % strain. In Fig. 3(a), we can see that the Li diffusion barrier increases under compressive strain while decreases under tensile strain. As the strain changes from −10% to 20 %, the energy barrier drops from 0.61 eV to 0.14 eV, which indicates that the Li diffusion coefficient becomes larger/smaller with the interlayer spacing of BLG elongated/compressed. Similar results were also found for interlayer Li diffusion barriers in stage-II compound in previous work.[17] It can be easy to understand this tuning effect of the Li diffusion barrier for layered structure by external out-of-plane strain. When the distance of two graphene layers increases/decreases under applied strain, the bonding effect between the diffusion Li atom and its adjacent carbon atoms becomes weaker/stronger, and thus the diffusion transition energy, i.e. energy barrier, decreases/increases. In fact, this phenomenon was also verified in other layered electrode materials from the first-principles calculations for uniform strain modulation.[18-20] Hao et al. summarized that the activation energy (diffusion barrier) of Li diffusion can be approximately expressed by a linear relationship with strain[66]

$$E_B = A_0 + A_1 \varepsilon \tag{1}$$

where $A_0$ and $A_1$ are material parameters. In this study, $\varepsilon$ is considered as the strain in the out-of-plane direction for layered materials in LIBs. Based on the above discussion, $A_1$ should be negative, representing the modulation effect on Li diffusion by uniform strain. The $A_0$ indicates the intrinsic Li diffusion barrier of materials.

Next, we explore the flexo-diffusion effect, induced by strain gradient. To compare the effects of uniform strain and strain gradient on diffusion, we calculate the energy barrier (from site A to site B) under various applied uniform strains and strain gradients, as shown in Fig. 3(b), and corresponding *x*-axis is aligning according to the schematic configures in it. The energy barrier significantly decreases/increases when a positive/negative strain gradient is applied. This trend is the same as the case of uniform strain, because the magnitude of strain in the positive/negative strain gradient field is also positive/negative. However, it is worth noting that the calculated diffusion barriers in strain gradient fields are near the $x = 0$ position of the supercell, where the magnitude of strain is much smaller than that of corresponding uniform strain case. Nevertheless, we can still explicitly find that the energy barrier variation of strain gradient is larger than that of uniform strain in Fig. 3(b). In addition, the diffusion barrier for the 5 % mid-strain case of strain gradient (155 meV) is almost the same as that of the 20 % uniform strain case (143 meV), which indicates the strain gradient is much more efficient than the uniform strain to tune the diffusion barrier. Meanwhile, the lattice distortion is much smaller in the 5 % mid-strain gradient case than that of 20 % uniform strain case, indicating the strain gradient can tune the barrier energy more efficiently than uniform strain with little lattice distortion, which is beneficial to the LIBs applications.

The temperature-dependent transition rate can be evaluated by the Arrhenius equation, from which the diffusion constant ($D$) of Li follows[67]

$$D = D_0 \exp\left(\frac{-E_B}{k_B T}\right) \qquad (2)$$

where $D_0$ is the proportionality coefficient, $E_B$ is the activation energy, $k_B$ is the Boltzmann's constant and $T$ is absolute temperature set to 300 K. According to equation (2), the diffusion mobility of Li along the positive strain gradient direction of 5 % mid-strain is about $1.6 \times 10^2$ times faster than that of pristine BLG at room temperature, the corresponding magnitude of strain gradient is about $8 \times 10^7$ m$^{-1}$ (comparable to the strain gradient in other nanostructures[30,31]). In general, effectively tuning Li diffusion mobility usually requires large strain, which is difficult to apply and it may also induce the material structure failure or even damage. Here, we find that a strain gradient induced by a small magnitude strain can modulate the diffusion properties, which is much more effective than uniform strain. The results suggest that lattice strain gradient can have a remarkable and efficient effect on the rate performance of batteries, with a significant increase in the ionic conductivity but only a slight change of the material structure at nanoscale. Therefore, extremely high-rate capability and high-efficiency modulation are expected for strain gradient modulated graphene-based or other 2D materials based LIBs. On the other hand, for the -8 × 10$^7$ m$^{-1}$ negative strain gradient case (-5 % mid-strain), the diffusion mobility of Li is ~ 10$^{-5}$ times smaller than that of pristine, indicating an extremely strong inhibition effect on Li diffusion. The negative strain gradient offers a significant suppression effect on diffusion, which suggests that this situation should be avoided in the flexible electronic devices in order to make LIBs work effectively. Besides, the huge distinction between positive/negative strain gradient effects on diffusion may introduce novel structural design, such as switch based on force sensor, to layered solid electrolytes, layered electrode materials even all-solid-state batteries.

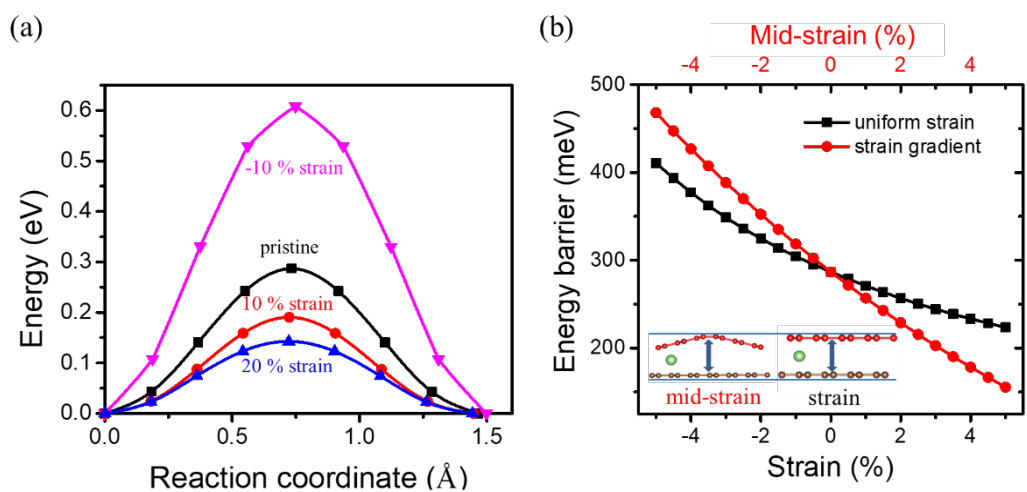

Fig. 3 (a) The energy profiles for Li pathways in four cases with uniform strain: pristine (0%), ±10 % and 20 %. (b) The variation of energy barrier of graphene as a function of uniform strain or mid-strain in strain gradient field. All the diffusion paths in the calculations are from site A to site B.

In order to study the mechanisms of flexo-diffusion effect and the reason for the difference of modulation effect between uniform strain and strain gradient, we calculate the potential energy profiles of Li in BLG along the diffusion path with five unit-paths, as shown in Fig. 4(a). Three conditions are considered for the strain fields: pristine, 5 % uniform strain and strain gradient with 5% mid-strain. Generally, the Li diffusion barrier is determined by the potential energy surface along the diffusion pathway, which is related to both the crystalline structure and the local atomic species of the compound. When the strain field is uniform (including pristine BLG), the Li atom experiences a periodic potential field established by the layered carbon framework, and thus the energy variation in the diffusion path is also periodic. When a positive strain gradient is introduced, the symmetry of the lattice is broken, and the Li potential energies of two adjacent stable positions are changed, as shown in Fig. 4(b). Due to the lower Li potential energy at the site with a larger strain, the asymmetry decreases the Li potential energy in the whole diffusion path. As a result, the peak of the Li potential energy of each unit path, i.e. diffusion barrier, in the diffusion path significantly decreases, thereby increasing the conductivity. In contrast, when Li atom diffuses along the negative strain gradient direction, the next stable position is under a lower strain and with a higher potential energy, and thus the energy barrier will increase. This is explicitly shown in the green curve in Fig. 4(b) from the right to the left, which indicates the negative strain gradient case. Our results demonstrate that the flexo-diffusion effect originates from the symmetry breaking of the potential energy field in crystal lattice due to the strain gradient.

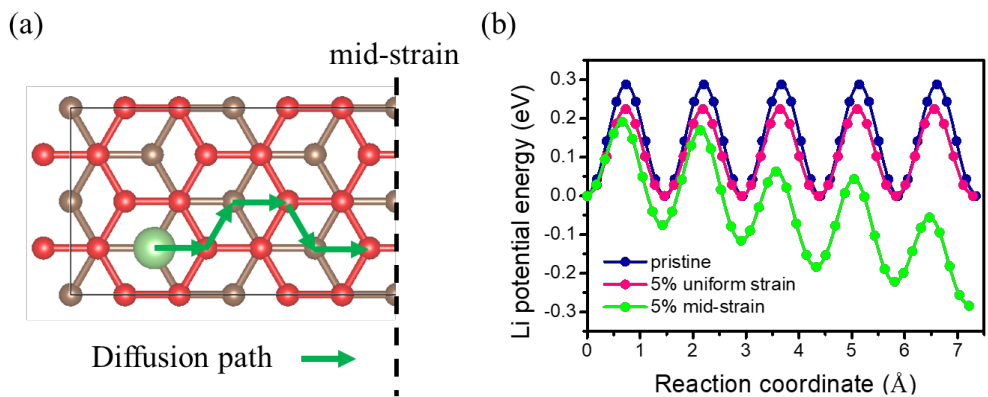

Fig. 4 (a) The left half part of 4 × 2 × 1 supercell and the calculation diffusion unit-paths hopping between adjacent HT sites. The red and brown represent the top and bottom carbon atomic layers, respectively. (b) The energy profiles of Li atom along the diffusion path in different strain fields. The magnitude of strain gradient for strain gradient configuration is about $3 \times 10^7$ m$^{-1}$.

To further explore the flexo-diffusion effect, we calculate energy barriers with additional uniform strain or diffusion path deviating from the $x$ direction. Firstly, to study the modulation effect of uniform strain on BLG already in strain gradient field, we apply additional uniform strain $\varepsilon_0$ to the strain gradient case, i.e. the interlayer spacing following $d(x) = d_0 \times (1 + \varepsilon_0 + \eta x)$. Fig. 5(a) shows the change of barrier with additional uniform strain based on ±5 % mid-strain configurations. It is observed that the linear strain modulation effect is also satisfied for strain gradient structures. Based on results in Fig. 3(b) and 5(a), the strain gradient effect could also be assumed as linear effect, superposition on the original strain effect. The influence of additional uniform strain on the -5 % mid-strain case is more obvious than that of the 5 % mid-strain case, due to the lager magnitude of energy barrier under negative strain. Secondly, it is expected that the Li diffusion deviating from strain gradient direction ($x$ direction, deviating angle is $\theta$) will also affect the energy barrier. The energy barriers of path-2 (from site B to site C in Fig. 2(a), $\theta = 60°$) under different strain fields are calculated, and are compared with that of path-1 (from site A to site B in Fig. 2(a), $\theta = 0$), as shown in Fig. 5(b). The flexo-diffusion effect for path-2 with $\theta = 60°$ is significantly weaker than that of path-1 with $\theta = 0$. This shows that the flexo-diffusion effect also depends on the deviating angle between unit diffusion direction and strain gradient direction, in which the flexo-diffusion effect achieves the maximum when $\theta = 0$ (enhancement for positive strain gradient) or $\theta = 180°$ (resistance for negative strain gradient). Therefore, there should be an angle correction $f(\theta)$, included in the relationship between energy barrier and strain gradient. Then Li diffusion energy barrier in layered compounds for a certain configure can be given in equation (3) which is further extended from equation (1) by introducing flexo-diffusion effect:

$$E_B = A_0 + A_1\varepsilon + A_2 \frac{d\varepsilon}{dx} f(\theta) \tag{3}$$

where $A_0$ is intrinsic Li diffusion barrier, $A_1$ is the response of energy barrier to uniform strain, $A_2$ is the response of energy barrier to strain gradient, $\frac{d\varepsilon}{dx}$ is the strain gradient in the $x$ direction. Based on our results and discussion, the parameter $A_2$ should also be negative. Though we don't know the exact function form of $f(\theta)$, we can take its value to 1 for the case of $\theta = 0$ for studying the diffusion properties along strain gradient direction. It will be beneficial to obtain the value of material parameters $A_0$, $A_1$ and $A_2$ for the accurate regulated response of strain for practical applications. From our DFT results, it is easy to fit the material parameters $A_0$, $A_1$ and $A_2$ for the case of $\theta = 0$, as $A_0$ = 0.29 eV, $A_1$ = -1.7 eV and $A_2$ = -1.9 × $10^{-9}$ eV·m. For most layered electrode materials and various diffusion atoms, these three parameters can also be calculated with similar DFT calculations, which will be of importance for the strain engineering and material genetic engineering in batteries.

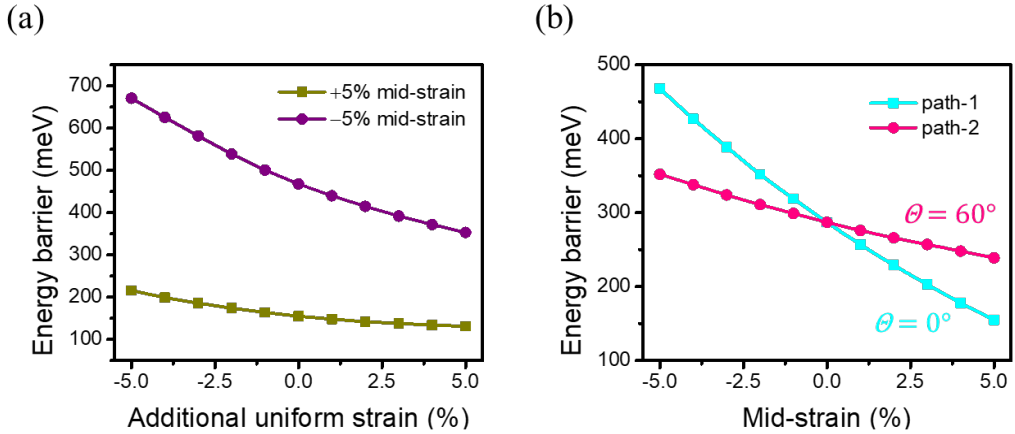

Fig. 5 (a) The change of energy barrier under the combination of uniform strain and strain gradient. The diffusion path is from site A to site B. The corresponding magnitude of strain gradient is about ±8 × $10^7$ m$^{-1}$ for the ±5 % mid-strain cases. (b) The effect of diffusion angle on the energy barrier, path-1 is from site A to site B ($\theta = 0$) and path-2 is from site B to site C ($\theta = 60°$), site A, B and C are shown in Fig. 2.

The strain and strain gradient can tune the diffusion barrier, which can help to design effective LIBs with mechanical strain. Because of the negative $A_1$, tensile strain can reduce the diffusion barrier, leading to an increase in conductivity. Nevertheless, applying tensile strain in the out-of-plane direction of materials of LIBs may be challenge in the applications. Instead, compressive strain is a better choice for the applications. However, compressive strain increases the diffusion barrier and causes the reduction of the conductivity, which is harmful for the LIBs. Thanks to flexo-diffusion effect, the conductivity can be still enhanced with the concentrated compressive stress, which can induce the positive strain gradient. The reason is that the enhancement of positive strain gradient to the conductivity overcomes the reduction effect of compressive strain. Therefore, concentrated compressive stress, which is applicable in the applications, could enhance the performance of LIBs with the flexo-diffusion effect. The detailed discussion can be found in the Supplementary Note 2.

It is noted that the gradient of some physical fields could drive oriented motions, such as thermal gradients on graphene to drive nanoflake motion,[68] chemical gradients on graphene to drive droplet motion[69] and proton-gradient-driven oriented motion of nanodiamonds grafted to graphene.[70] Thus, we also examine the motion of Li atom inside BLG in different strain fields by AIMD at 500 K. Fig. 6 shows the displacement change of Li atom with time in different strain fields: pristine (0 %), 3 %, 5 % and 7 % mid-strain. The displacement is defined as

$$d = \sqrt{(x-x_0)^2 + (y-y_0)^2 + (z-z_0)^2} \qquad (4)$$

where $(x_0, y_0, z_0)$ and $(x, y, z)$ are the initial and current Li coordinates, respectively. When the strain gradient is zero or small, Li atom only keeps thermal motion around the initial stable site A. However, as the magnitude of strain gradient increases, Li atom will gradually move away from the initial site A. Interestingly, the simulated motion direction is exactly along the diffusion path discussed in the previous sections. For the case of 5 % mid-strain, the simulated motion result is from site A to site B, and Li atom can't keep moving to site C due to the deviation from the strain gradient direction, which has a higher energy barrier and requires a larger strain gradient to overcome. For the case of 7 % mid-strain, the reason for that Li atom moves back to site A around 2 ps is due to thermal perturbation. Given that the energy barrier of path from site B to site C further deceases, the Li atom can gradually move from site A to site B to site C to site D. It shows that the strain gradient can drive oriented motion of Li atoms inside the BLG, the orientation contains two aspects: positive strain gradient direction and along the diffusion path. Moreover, the motion speed could be tuned by the magnitude of the strain gradient. Such tunability and phenomenon in graphene provide additional capabilities in device design for applications, such as self-charging technique and electromechanical sensors and actuators.

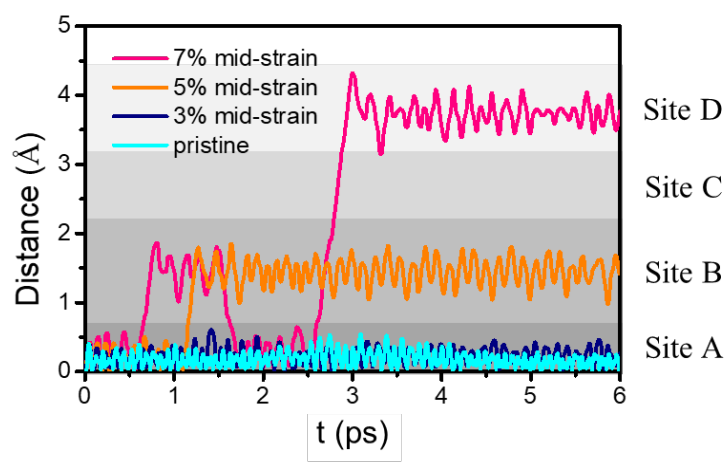

Fig. 6 The change of displacement and position of Li atom inside BLG with time under different strain fields simulated by AIMD at 500 K. Each interval (presented in different shades of gray) indicates the range of corresponding site position.

## 4. Conclusions

We propose the flexo-diffusion effect and investigate the diffusion properties of Li atom inside BLG with uniform strain and gradient strain fields from DFT calculations. It is found that the Li diffusion barrier in BLG is very sensitive to the strain gradient applied to the lattice, which is termed as flexo-diffusion effect. The Li diffusion barrier decreases/increases substantially upon the positive/negative strain gradient applied along the in-plane direction, leading to the change of Li diffusion coefficient in several orders of magnitude at 300 K. Moreover, the effect of regulation for strain gradient is more significant than that of uniform strain field. We also investigate the Li potential energy distribution in BLG in different strain fields, which shows that the novel effect induced by strain gradient is attributed to the asymmetric potential energy distribution for the Li diffusion. The asymmetric lattice structure also provides a driving force for the oriented motions of Li atoms, and the motion speed could be tuned by the magnitude of strain gradient. Furthermore, graphene is used as a typical 2D layered electrode material, and we expect that the flexo-diffusion effect could also be applicable to other layered materials where Li atoms store and diffuse between interlayers. The flexo-diffusion effect may extend present LIBs technologies by introducing the novel strain gradient engineering.

## Conflicts of interest

There are no conflicts to declare.

## Acknowledgements

The research support of the National Natural Science Foundation of China (No. 11572040, No. 11521202, and No. 11804023) , the National Key R&D Program of China (2016YFB0700600) and the Foundation for Innovative Research Groups, the Thousand Young Talents Program of China and National Key Research and Development Program of China (Grant No. 2016YFB0402700) are gratefully acknowledged.

## Notes and references